\documentclass[twocolumn,prb,showkeys,preprintnumbers]{revtex4}

\usepackage{graphicx}
\usepackage{amssymb}
\usepackage{xspace}
\usepackage{amsmath}
\usepackage {subfigure}
\usepackage{fancyhdr}

\newcommand{\bvec}[1]{\ensuremath{\mathbf{#1}}}
\newcommand{\bm}[1]{\ensuremath{\mathbf{#1}}}
\newcommand{\bhat}[1]{\hat{\ensuremath{\mathbf{#1}}}}

\newcommand{\bq}{{\bf q}}

\newcommand{\bkin}{{\bf k}_{\mathrm{in}}}
\newcommand{\bkout}{{\bf k}_{\mathrm{out}}}

\newcommand{\FT}{{\cal F}}

\begin{document}
\title{Compressive auto-indexing in femtosecond nanocrystallography}

\author{Filipe R. N. C. Maia}
\affiliation{Lawrence Berkeley National Laboratory, 1 Cyclotron Road, Berkeley, CA 94720, USA}
\author{Chao Yang}
\affiliation{Lawrence Berkeley National Laboratory, 1 Cyclotron Road, Berkeley, CA 94720, USA}
\author{Stefano Marchesini}
\email{smarchesini@lbl.gov}
\affiliation{Lawrence Berkeley National Laboratory, 1 Cyclotron Road, Berkeley, CA 94720, USA}




\begin{abstract}
  Ultrafast nanocrystallography has the potential to revolutionize
  biology by enabling structural elucidation of proteins for which it
  is possible to grow crystals with 10 or fewer unit cells on the
  side.  The success of nanocrystallography depends on robust
  orientation-determination procedures that allow us to average
  diffraction data from multiple nanocrystals to produce a three
  dimensional (3D) diffraction data volume with a high signal-to-noise
  ratio.  Such a 3D diffraction volume can then be phased using
  standard crystallographic techniques.  ``Indexing'' algorithms used
  in crystallography enable orientation determination of diffraction
  data from a single crystal when a relatively large number of
  reflections are recorded.  Here we show that it is possible to
  obtain the exact lattice geometry from a smaller number of
  measurements than standard approaches using a basis pursuit solver.
\end{abstract}

\keywords{crystallography; indexing; compressive sensing}
\eprint{LBNL-4008E}
\maketitle

%
\fancyhead[RO]{LBNL-4008E}
\fancyhead[LO]{\slshape \leftmark}

\pagestyle{fancy}





\section{Introduction\label{Introduction}}

X-ray crystallography is currently the leading method for atomic
resolution imaging of macromolecules. Third generation synchrotron
sources permit successful structure solution from crystals 5 microns
in size or greater \cite{Holton:xh5015}. Obtaining sufficiently large
crystals is currently an important stumbling block in structure
determination.

The Linac Coherent Light Source (LCLS) recently began operation
\cite{LCLS} at the SLAC National Accelerator Laboratory in Palo Alto,
California, using energetic electrons from a linear accelerator to
produce coherent x-rays with an instrument called a free electron
laser (FEL).  Free Electron Laser sources produce pulses of light that
are over 9 orders of magnitude brighter than current third generation
synchrotron sources \cite{FLASH}. Several other x-ray laser sources of
this type are being built or planned worldwide \cite{Shintake,
  Altarelli}.

The high number of photons incident on a specimen are expected to
produce measurable diffraction patterns from nanocrystals perhaps even
down to a single period (single molecule) \cite{chapman:natmat09},
enabling high resolution structure elucidation of systems which can
only be crystallized into very small crystals that are not suitable
for conventional crystallography. Even for larger crystals, the short
pulses can circumvent the radiation damage problem \cite{Neutze,
  Chapman} which limits the resolution of many sensitive crystals.

In such an experiment, two-dimensional (2D) diffraction images of
randomly oriented nanocrystal of the same type can be captured within
an exposure time of a few femtoseconds. These images can then be used to deduce the
3D structure of the molecule. To see the structure in 3-D, one has to
merge the data from all these individual nanocrystals, whose
orientations are not known.

Femtosecond nanocrystallography brings new challenges to data
processing \cite{kirian}.  One problem is that the orientation of each
diffraction image obtained is unknown. Another problem is that a
single snapshot of the crystal diffraction pattern may contain very
few reflections.  In traditional crystallography, a small angular
range of integration ensures that many Bragg reflections are recorded
while ensuring that overlaps are minimized. This is not possible with
ultrafast x-ray pulses. The relentless improvements of these light
sources (beam energy, beam divergence and wavelength) will further
exacerbate the problem.

These new difficulties make indexing such patterns a hard problem for
existing crystallographic software.


\section{Structure Determination from Crystal Diffraction}

\begin{figure}[htbp]
  \includegraphics[width=0.45\textwidth]{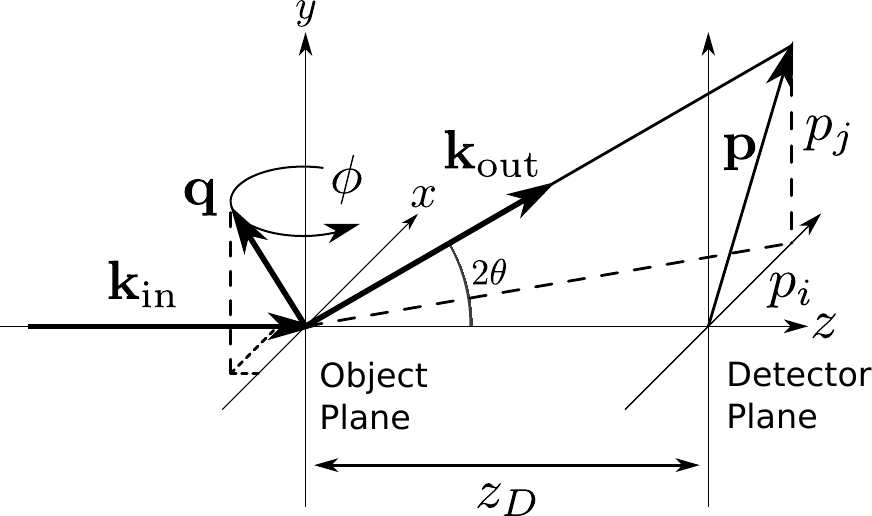}
  \caption{Scattering geometry for coherent x-ray diffraction imaging.}
  \label{fig:geometry}
\end{figure}

 The diffracted photon flux $I$ (photons/pulse/ pixel) produced by a
crystallite is given by
\begin{eqnarray} 
\label{eq:1}
I(\bm{q})&=&J_o r_e^2 P \Delta \Omega 
\left | F(R_\phi \bm{ q})\right | ^2 \cdot \\ 
\nonumber 
&\cdot& 
\left | \sum_{h,k,l}S\left ( 
R_\phi\bm{ q}- (h \bhat{h}+k \bhat{k}+l
\bhat{l}) \right )\right | ^2,
\end{eqnarray}

where $F(\bm{q})$ is the continuous scattering from one unit cell
(molecule), $R_\phi$ is the 3D rotation matrix of the unknown object
orientation, $\bm{q}$ a 3D vector that relates the Bragg
``reflection'' on a two dimensional detector to a point in a 3D
Fourier space, $J_o$ is the incident photon flux density
(photons/pulse/area), $r^2_e$ is the electron cross section, $P$ is a
polarization factor, $\Delta \Omega$ is the solid angle subtended by a
detector pixel at the sample, the $(h,k,l)$ integer values are called
Miller indices, $(\bhat{h},\bhat{k},\bhat{l})$ identify the Bravais
lattice characteristic of the crystal periodic structure, and $S$ is
the shape transform of the crystallite finite dimensions. For large
crystals, $S$ is simply a Dirac $\delta$-function.  For a crystal made
of a few unit cells, $S$ is broadened and may introduce an error in
the location of the reflection.


Each set of pixel coordinates on a detector placed at a distance $z_D$ from
the sample is $\bm{p}_{i,j,z_D} = ( p_i \bhat{x} + p_j \bhat{y}+z_D \bhat{z})$
and corresponds to a value of $\bq$ in the 3D reciprocal space according to
the geometric description of elastic scattering shown in
Figure~\ref{fig:geometry}. 

 In this figure, $\bkin$ and $\bkout$ are the incident and scattered wave
vectors that satisfy $|\bkin | = |\bkout | =k= 1/\lambda$, where
$\lambda$ is the wavelength of the x-ray.  The direction of $\bkin$
and $\bkout$ are the same as the direction of the incident beam
$\bkin=k (0,0,1)$ and the outgoing beam 
$\bkout= k \bhat{p}_{i,j,z_D}$. 
  The coordinates of a lattice point $\bq_{i,j}$ corresponding
to $p_{i,j, z_D}$ satisfy
\begin{eqnarray}
\nonumber
  \bvec{q}_{i,j} &=& \bkout -\bkin\,,\\
  &=& 
  \label{eq:CCD_q_ij}
  \frac{1}{\lambda} \,
  \left ( \tfrac{(p_i,p_j, z_D)}
               {
                 \sqrt{p_i^2+p_j^2+z_D^2} 
		} -
    (0,0,1) \right ).
\end{eqnarray}
The end point of the vector $\bq$ lies on a 2D surface called 
the Ewald sphere.  This spherical surface of radius $k$
intersects the origin ($\bq=0$ when $\bm{p}=(0,0,z_D)$), and is
centered at $(0,0,-k)$ (while the origin of $p_{i,j,z_D}$ is at the
sample).


In traditional crystallography, the most common data collection method
is the rotation method in which the diffraction image is collected while
rotating the sample, i.e.  varying $R_\phi$ in Eq. \ref{eq:1}.


A small angular range of integration ensures that all Bragg
reflections are recorded while overlaps are minimized.
The strength (measured intensity) and orientation of each Bragg
reflection is estimated from the diffraction geometry (including
source divergence, bandwidth, pixel size and angular average). 


In x-ray crystallography, the term {\em indexing} refers to the task
of assigning the measured Bragg peaks to the discrete locations
$(h,k,l)$ of a periodic lattice.  Auto-indexing uses the position of
these peaks to deduce the shape $(\bhat{h},\bhat{k},\bhat{l})$ and
orientation ($R_\phi$) of the lattice, and to identify the lattice
coordinates $(h,k,l)$ of each measured peak.

It is accomplished in several steps.

\begin{itemize}

\item For the purpose of autoindexing, one can simply assign the value
  of 1 to $I(\bq_{i,j})$ for every peak above a noise threshold.  As a
  result, one obtains a 3D map $b(\bq)$ in the reciprocal space that
  contains the values of either 1 or 0.  Note that $b(\bq)$ is only
  affected by the content of a unit cell when $|F(\bm{\bq})|$ is so
  small that the reflection is not detected. 
Assuming $S(\bq)\simeq \delta(\bq)$, Eq. (\ref{eq:1}) becomes:
\begin{equation} 
\label{eq:1a}
b(\bm{q})\simeq \sum_{h,k,l}\delta\left ( 
R_\phi\bm{ q}- (h \bhat{h}+k \bhat{k}+l
\bhat{l}) \right ),
\end{equation}

\item Some type of computational analysis is performed on the 3D map
  to ascertain the orientation and the unit cell parameters of the
  crystal ($R_\phi,\bhat{h},\bhat{k},\bhat{l}$).  The analysis
  typically proceeds by making use of Fourier transform and peak
  searches.  An efficient algorithm that uses many 1D Fourier
  transform was proposed in \cite{steller, rossmann}. It is used in
  many existing autoindexing software packages such as MOSFLM
  \cite{mosflm}. We will provide details of these algorithms in the
  next section, as this problem will be the focus of our paper.

\item Once the lattice vectors and orientation are determined, the
  lattice coordinates that overlap with the Ewald sphere will provide
  the index of a reflection. Multiple solutions due to mirror
  symmetries of the lattice are generated. These solutions can be
  distinguished using the measured intensities.

\end{itemize}

Once the orientation and the unit cell parameters associated with
a crystal has been determined, one may then proceed to estimate
the intensities of the crystal, from the diffraction geometry (including
source divergence, bandwidth, pixel size and angular average).
Finally, a phase retrieval algorithm is used to recover the phase of
the Fourier transform and subsequently the 3D density map of the
crystal.

For the purpose of this paper, we will not discuss the issues of
structure factor determination or the phase retrieval problem.
Instead, we will focus on the second item of the autoindexing problem, how to
determine the lattice parameters and orientation.

Multiple solutions due to symmetries of the lattice (but not of the
crystal) will still have to be sorted out using measured
intensities. In this paper we do not address this problem, which
presents another challenge when attempting to merge many thousand of
low-signal snapshots.

\section{Real space autoindexing}


Most autoindexing algorithms search for peaks in real space, by
applying some form of 3D Fourier transform of the
binary reciprocal space map $b(\bq)$. 

If the region of $\bq-$space that was measured is large, its 3D FT
will provide the real space lattice.  A simple numerical thresholding
may reveal the positions of the 3D lattice points in real space. They
can subsequently be used to determine the unit cell parameters,
crystal orientation and type. 

The binary 3D mask $M(\bq)$ that defines the region of $\bq-$space that was
measured ($M(\bq)$=1 if $\bq$ was measured, =0 otherwise) can be viewed as a
3-dimensional optical transfer function.  The 3D Fourier transform of
the binary mask forms a point spread function which convolves the real
space lattice.  If the PSF is larger than the lattice
spacing, these cannot be resolved.

As we will show in section~\ref{example}, when the number of measured
Bragg peaks is less than 10, the real space lattice points cannot be
easily distinguished from the rest of the sampled voxels based on the
intensity of the inverse 3D Fourier transform.

The use of a 3D Fourier transform around the origin for
indexing a diffraction pattern was suggested over two decades ago
\cite{bricogne}. A similar approach appears to have been incorporated
in the program DENZO, which has been distributed as part of the
diffraction-image processing suite HKL \cite{otwinowski}.  A
3D FFT has been used to index diffraction images by
calculating a Patterson function from a set of reflections which have
all been assigned unit intensity \cite{Campbell}.  
Efficient software implementations avoid
the use of a full 3D Fourier transform, by using the
Fourier projection-slice theorem \cite{natterer}, calculating 1D
sections of the 3D FTs by a series of projections and
1-dimensional FFTs \cite{rossmann} \cite{powell}.  Indexing software
such as MOSFLM \cite{steller}, LABELIT \cite{Sauter:dd5008} utilize
this approach.

The complexity of the projection-FFT approach is $m n \log n$, where
$m$ is the number of direction vectors that must be generated, and $n$
is the number of samples along the projected 1D intensity profile,
which is proportional to $N^{1/3}$, where $N$ is the total number of
sampled voxels contained in the crystal.  A typical value of $m$ is
between 5,000 and 20,000. Clearly, this method will not work well if
the number of Bragg points on a diffraction pattern is small.

Although the argument used in \cite{rossmann} for abandoning the full
3D FFT is the high cost for performing 3D FFTs of large crystals, this
is no longer a serious issue due to the rapid growth in the processing
speed and memory capacity of modern multi-core microprocessors.  At
the time of writing, a $512^3 $ 3D FFT takes about 0.15 seconds on a
GPU processor, while a $2048^3$ takes about 7 minutes on a machine
with sufficient memory. Furthermore, there are now algorithms that we may use
to take full advantage of the sparsity of the 3D reciprocal space map
\cite{lexing}, i.e., there are a few non-zeros in the 3D map
constructed from the Bragg reflections, and reduce the complexity of
the 3D FFT calculation from the standard $\mathcal{O}(N \log N)$ to
that of $\mathcal{O}(N^{2/3} \log N)$, where $N$ is the total number
of sampled voxels in the crystal.

\section{Recovering Real Space Lattice via L1 Minimization}
An alternative technique for retrieving the positions of the real (and
reciprocal) space lattice points associated with a crystal is to use
the recently developed compressive sensing methodology
\cite{donoho,crtmath06,crtieee06} and formulate the problem as an L1
minimization problem.

Let $x$ be a vector representation of the 3D density map of a crystal lattice 
to be determined in real space.  Similar to the 3D inverse Fourier
transform approach, we will use the magnitude of each component of $x$ to
determine whether the 3D voxel associated with that component is a real
space lattice point.

The vector $x$ is related to the diffraction measurement through the
following equation:
\begin{equation}
b_i = e_{j_i}^T \FT x, \ \ \mbox{for} \ \ i = 1, 2,..., 2m, \label{sensing}
\end{equation}
where $b_i$ is the $i$th pixel of the binary image corresponding to a
sampled 3D reciprocal space voxel that lies on the Ewald sphere, $m$ 
is the total number of voxels on the 
Ewald sphere, $\FT$ is the matrix representation of a
3D discrete Fourier transform, $j_i$ is the index of the voxel 
(in $x$) that lie on the Ewald sphere, and $e_{j_i}$ is the $j_i$th column 
of the identity matrix.  To ensure that $x$ is real, Friedel's conjugate 
symmetry is imposed on $b$.  Therefore, we have $2m$ equations in 
(\ref{sensing}) even though the number of samples on the Ewald sphere is
$m$.

Because the number of sampled voxels on a Ewald sphere is always far fewer 
than the number of reciprocal lattice points, the linear system defined
by (\ref{sensing}) is clearly underdetermined. Therefore, $x$ cannot
be recovered by simply solving (\ref{sensing}). However, because
the vector $x$ to be recovered is expected to be ``sparse'', i.e., it is
expected to have nonzero values at a subset of real space voxels, it 
follows from the recently developed compressive sensing theory 
\cite{crtmath06,crtieee06}, that we may be able to recover $x$ by solving the following convex minimization problem

\begin{equation}
\begin{array}{cc}
\min_{x} & \| x \|_1 \\
\mbox{such that} & M \FT x = b,
\end{array}
\label{l1min}
\end{equation}
where $M$ is an $2m \times n$ sparse ``sensing'' matrix that contains
$e_{j_i}^T$, $i=1,2,...,2m$, as its rows, $b$ is a vector
representation of the intensity values (0's and 1's) assigned to
voxels on the Ewald sphere (and its Friedel symmetric counterpart),
$\| \cdot \|_1$ denotes the $L_1$ norm of a vector.

 The equality constraint in (\ref{l1min}) can be relaxed to an inequality
 constraint of the form
 \[
 \| M \FT x -b \|_2 \leq \sigma,
 \]
 where $\|\cdot \|_2$ denotes the $L_2$-norm of a vector, for some small 
 constant $\sigma$ to allow imprecise measurements or noise
 in the data. The relaxed minimization problem is often known as 
 the basis pursuit denoising (BPDN) problem, and the original L1 minimization
 problem (\ref{l1min}) is also known as the basis pursuit (BP) problem.

\begin{figure*}[htbp]
\subfigure[Volume rendering of the reciprocal space lattice produced from the 3D Fourier transform of
     d).]
  {
     \label{subfig:2a}
\includegraphics[trim=1.8in 3.5in 1.6in 3.3in, clip, width=0.4\textwidth ]
{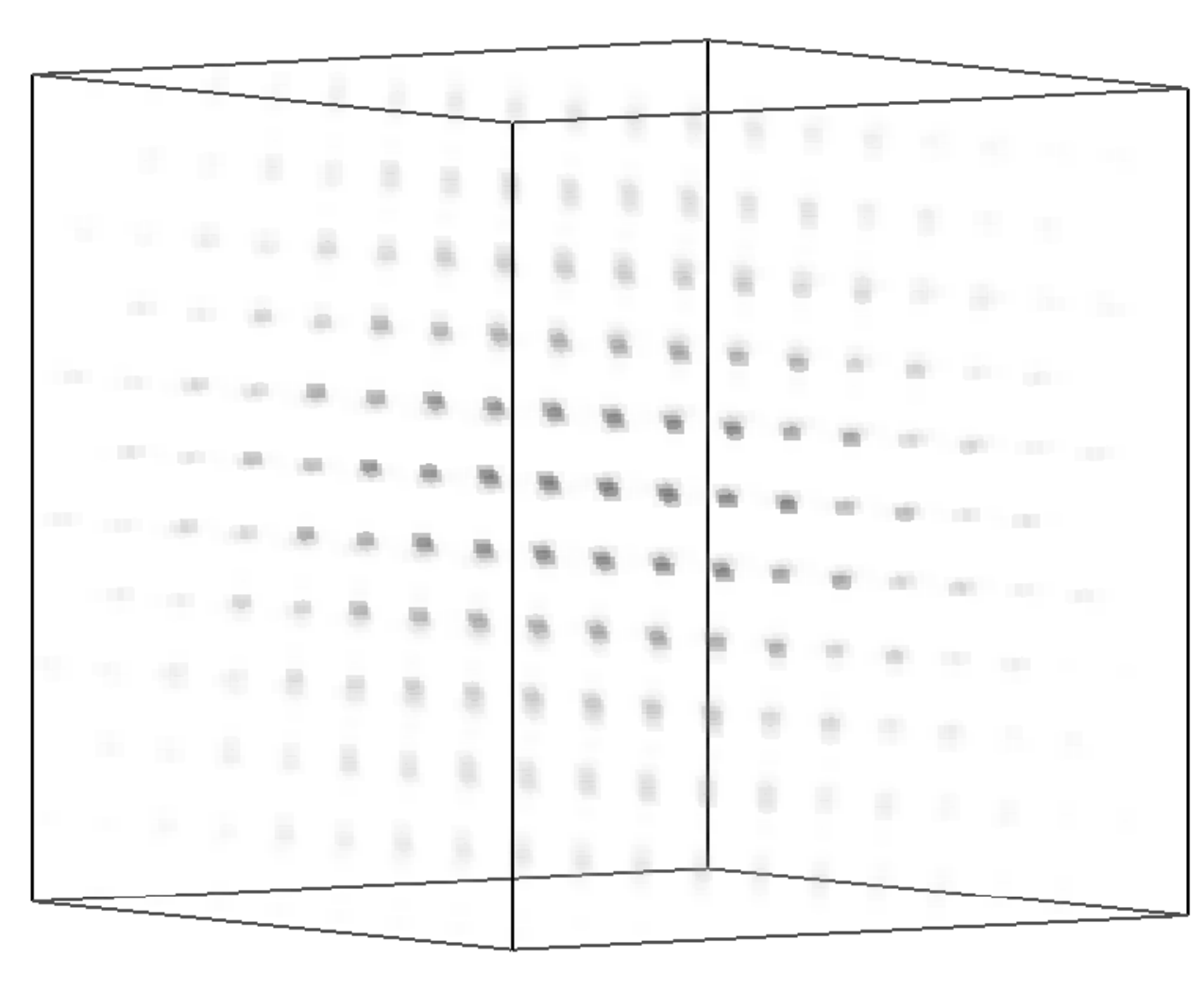}	
  }
\hfill
\subfigure[Surface representation of the Ewald
     sphere with the reflections that fall on it and marked as red
     dots.]
   {
     \label{subfig:2b}
\includegraphics[trim=1.3in 3.5in 1in 3.3in, clip, width=0.4\textwidth ]
{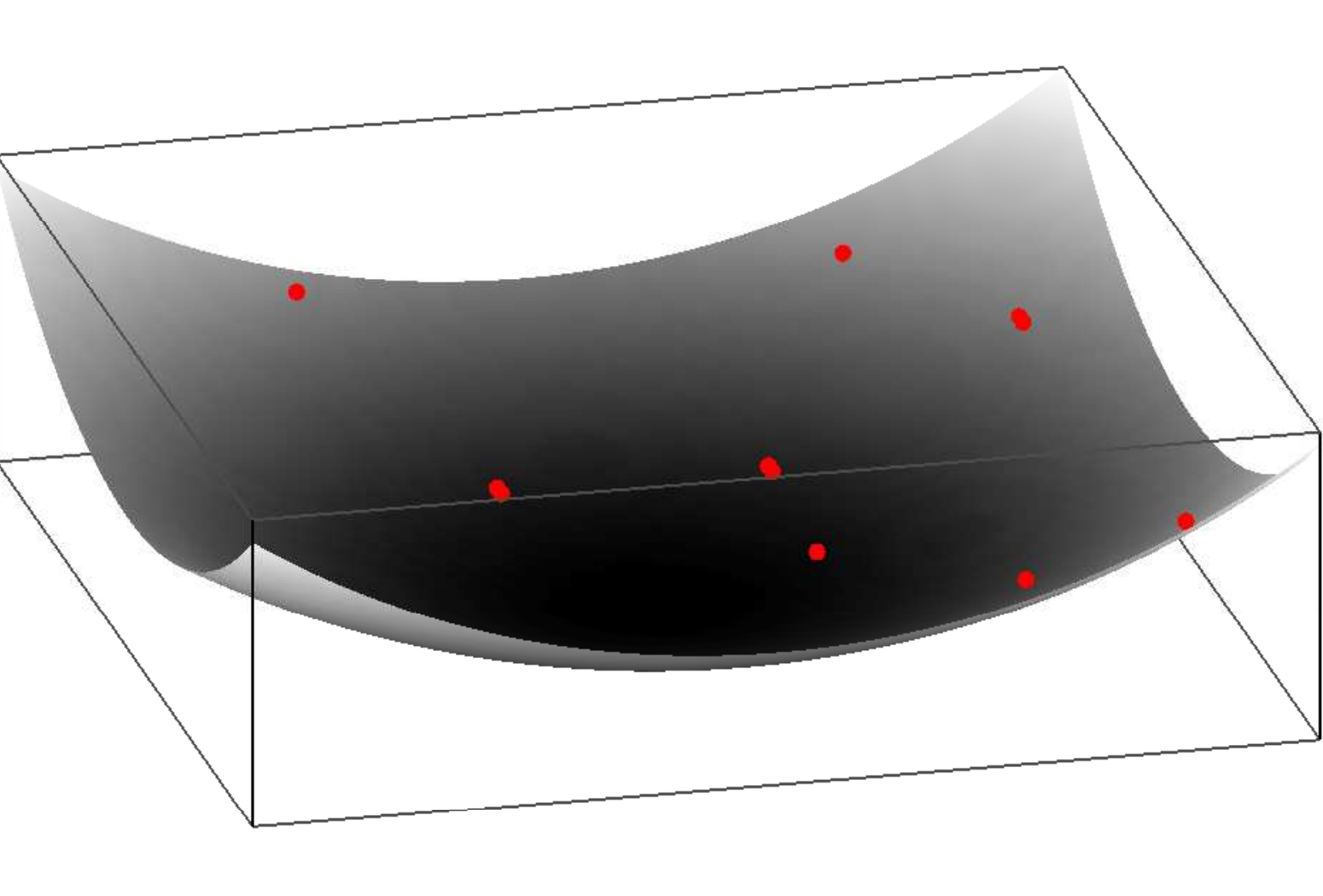}	
   }
\subfigure[``Observed'' diffraction pattern.]
    {
      \label{subfig:2c}
\includegraphics[trim=1.8in 3.5in 1.6in 3.3in, clip, width=0.4\textwidth ]
 {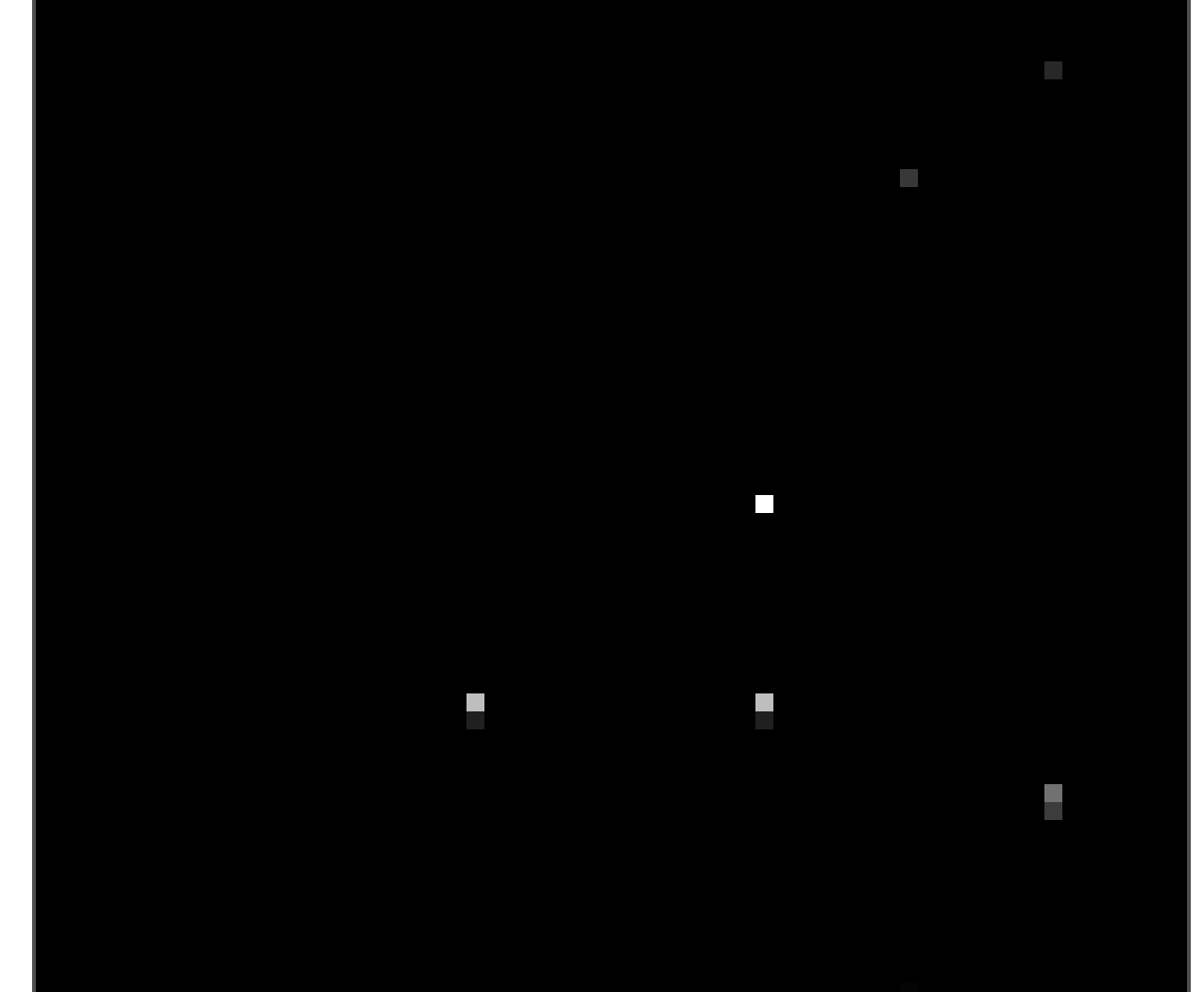}
    }
\hfill
\subfigure[Volume rendering of the real space volume showing the
      rotated crystal lattice.]
    {
      \label{subfig:2d}
\includegraphics[trim=1.8in 3.5in 1.6in 3.3in, clip, width=0.4\textwidth ]
{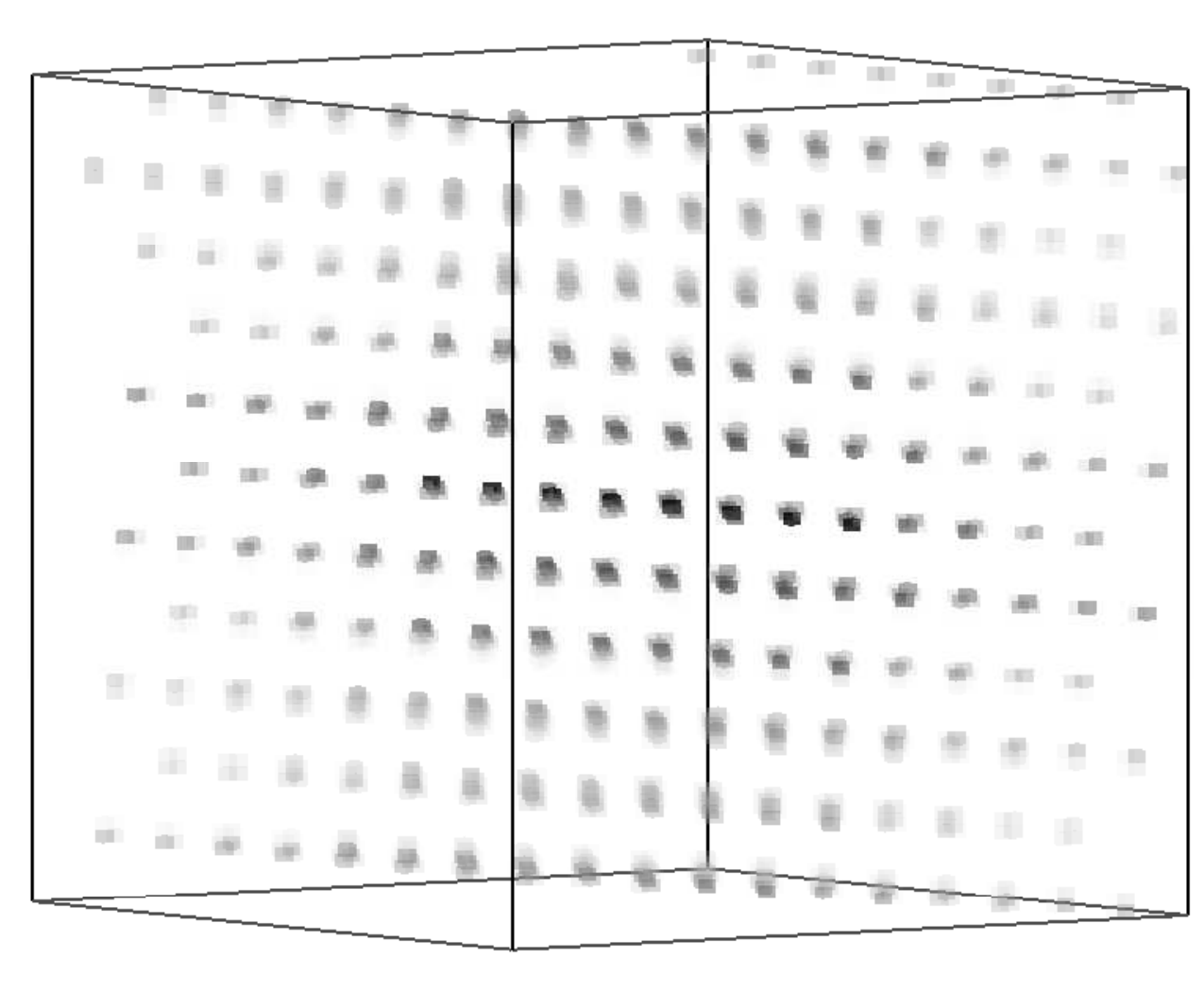}	
    }
\subfigure[Reconstructed real space volume using the 3D inverse Fourier
transform of the observed data.]
    {
      \label{subfig:2e}
\includegraphics[trim=1.8in 3.5in 1.6in 3.3in, clip, width=0.4\textwidth ]
{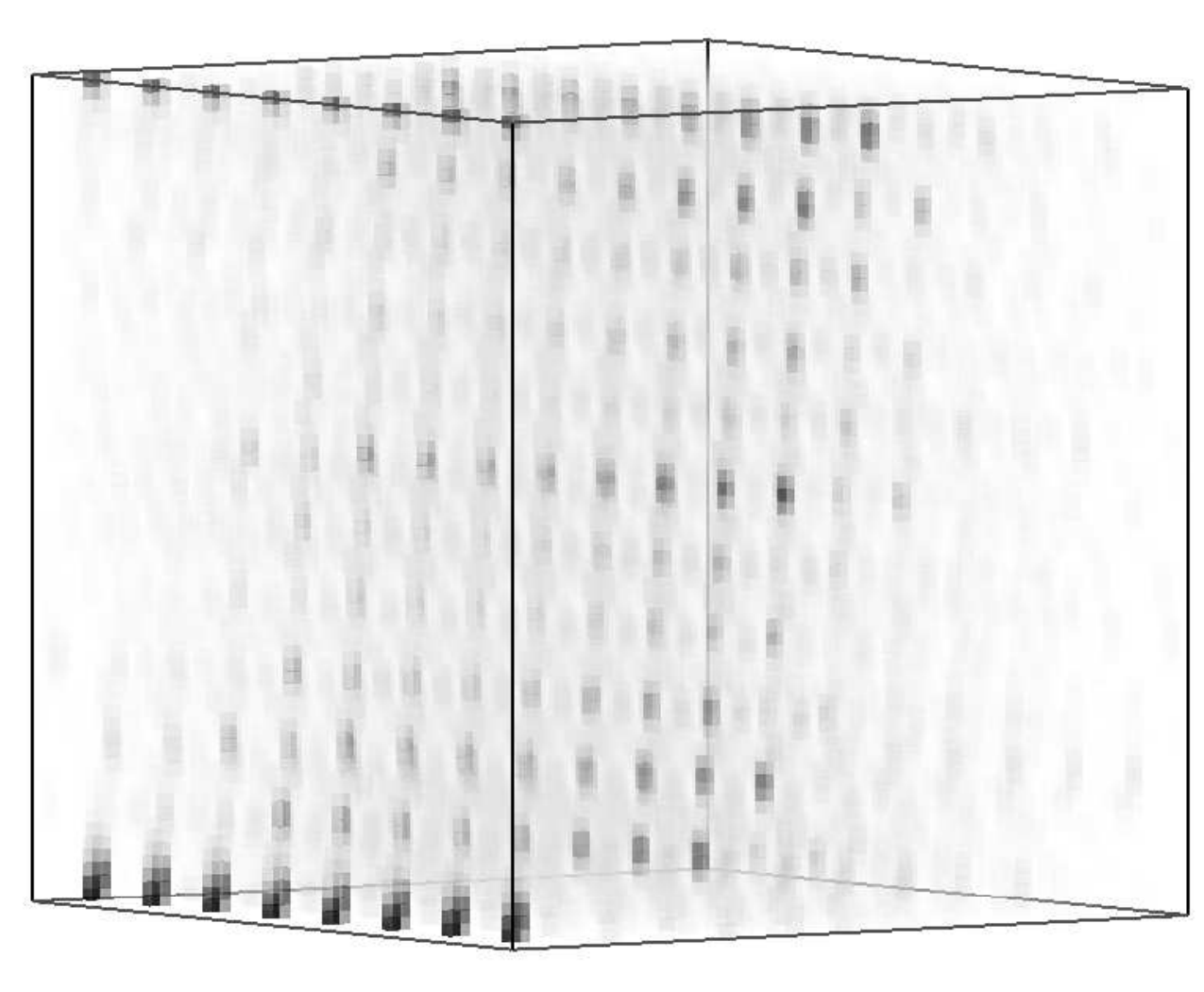}	
    }%
\hfill
\subfigure[Reconstructed real space volume using the L1 minimization method for 300 iterations.]
    {
      \label{subfig:2f}
\includegraphics[trim=1.8in 3.5in 1.6in 3.3in, clip, width=0.4\textwidth ]
{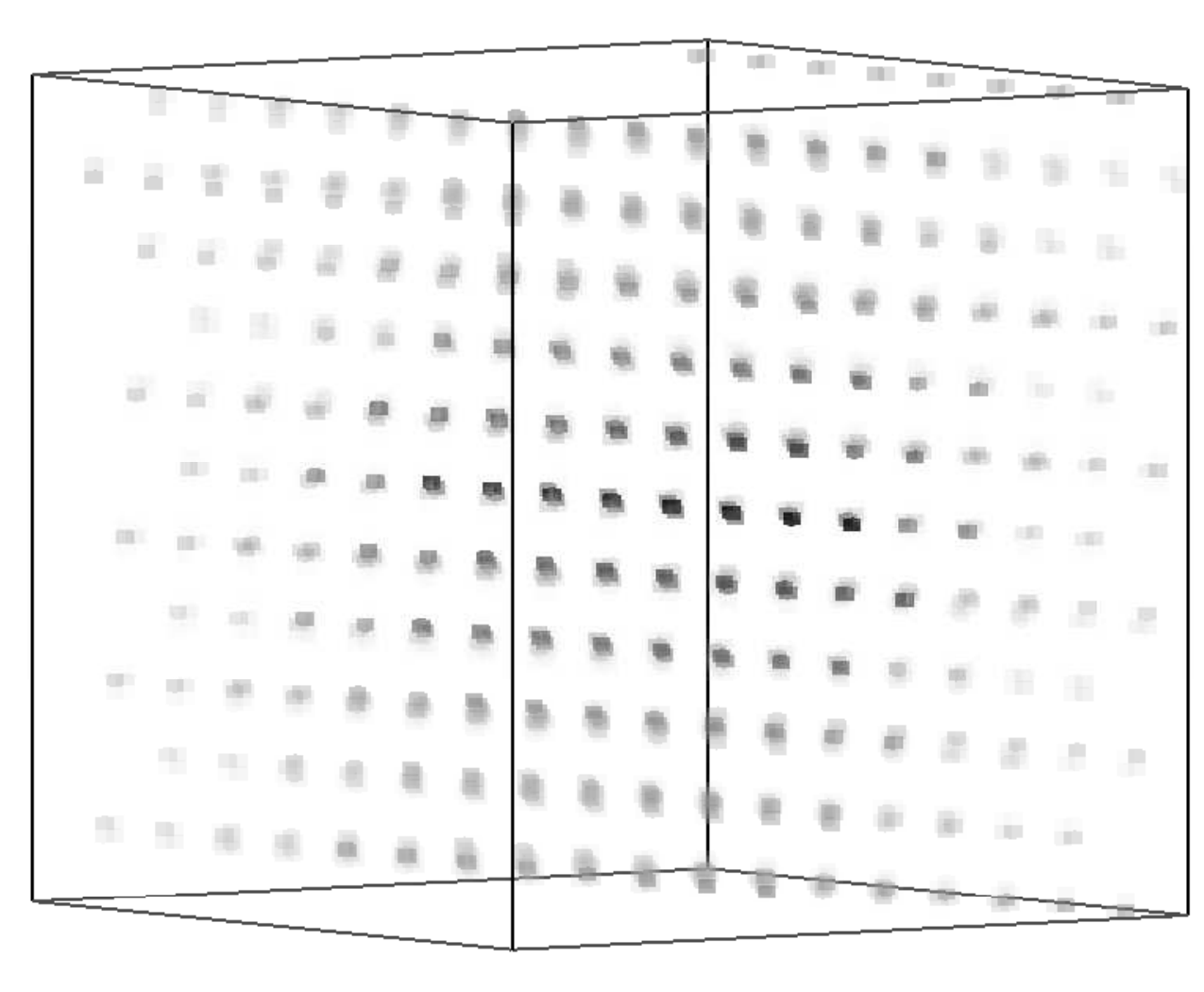}	
    }
    \caption{
Test problem: $64^3$  volume populated with a cubic lattice of period  $8 \times 8 \times 8$.  }
\label{fig:2}
\end{figure*}

\section{Algorithms for Solving the L1 Minimization Problem}
The L1 minimization problem (\ref{l1min}) and its BPDN relaxed form
can be solved in a number of ways. In the software package SPGL1
\cite{spgl1}, which we use to perform the numerical experiments shown
in the next section, the BPDN problem is reduced to a sequence of what
is known as the LASSO \cite{lasso} problems
\begin{equation}
\begin{array}{cc}
\min_{x} & \| M \FT x - b\|_2 \\
\mbox{such that} & \| x \|_1 \leq \tau,
\end{array}
\label{lasso}
\end{equation}
where $\tau$ is a parameter that is determined in an iterative process
that involves finding the root of nonlinear equation $\phi(\tau) =
\sigma$, where $\phi(\tau)$, which is the optimal value of the
objective function in (\ref{lasso}) for a given $\tau$, is known as
the Pareto curve.  The LASSO problem is solved by a spectral projected
gradient method \cite{birgin00,birgin03,dai} in SPGL1.

An alternative approach for solving the BPDN problem is to apply a
first-order method developed by Y. Nesterov
\cite{nesterov83,nesterov05} to solve (\ref{l1min}) directly. A
software package based on this approach is called NESTA \cite{nesta}.

The computational cost of both SPGL1 and NESTA is dominated by the
calculation of $\FT x$ and $\FT^{\ast}x$, i.e., 3D fast Fourier and inverse
Fourier transforms required in each iteration. Therefore, the overall
cost of an autoindexing scheme based on L1 minimization formulation of
the problem is higher compared to the existing approaches.  However,
as we will see in the next section, the advantage of the method is
that it can recover the real space 
lattice
points reliably even when only a few Bragg peaks can be identified on
a diffraction image.

%
%
%
%

 \begin{figure}[htbp]
   \centering 
\includegraphics[trim=.5in 2in 1in 2in, clip, width=0.45\textwidth ]
{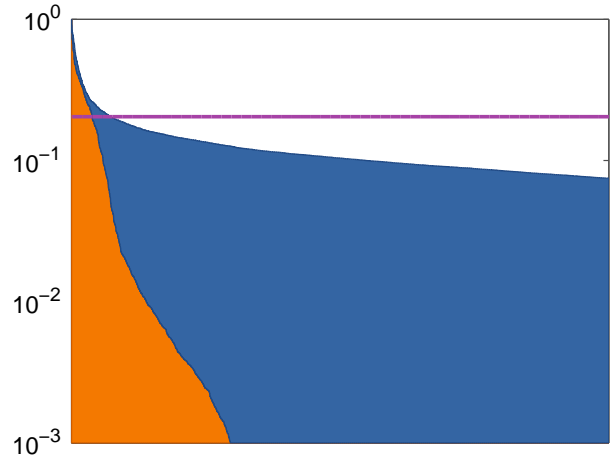}
   \caption{
\label{fig:3}
Normalized intensity values of the voxels reconstructed,
   sorted in descending order. The blue plot corresponds to the 3D
   Fourier transform method while the orange plot corresponds to the L1
   minimization method (after 300 iterations).}  
 \end{figure}

\section{Computational Experiment}
\label{example}
To test the algorithm we created a $64^3$ real space volume which was
then populated with a cubic lattice that contains $8 \times 8 \times
8$ voxels. A rotation was then applied to the lattice (fig.
\ref{subfig:2a}) and the result was Fourier transformed to generate
the 3D diffraction volume (fig. \ref{subfig:2d}). The size of the
problem was chosen for visualization purposes.

The simulated diffraction data was calculated by selecting those
voxels which are crossed by the Ewald sphere (fig. \ref{subfig:2b})
and projecting them onto the detector plane according to the geometry
shown in Figure~\ref{fig:geometry}. Figure \ref{subfig:2c} shows the
simulated 2D diffraction pattern. The detector plane is uniformly
sampled with $64 \times 64$ pixels, and we set the distance between
the crystal and the detector to 64 pixels.

We then tried to recover the real space lattice using two methods. In
the first method we simply took the inverse 3D FFT of the diffraction
volume in which the voxels that were not ``observed'' were set to
zero.  The intensity of the transformed volume is shown in Figure
\ref{subfig:2e}. In the second approach, we solve the L1 minimization
problem (\ref{l1min}) discussed in the previous section by using the
SPGL1 software provided by the authors of \cite{spgl1}.  The amplitude
of the reconstructed real space lattice $|x|$, after 300 iterations of
SPGL1 is shown in Figure~\ref{subfig:2f}. After only 10 iterations it
is already possible to see the unit cell positions clearly enough to
determine the crystal orientation and unit cell dimensions.


It is clear from Figures~\ref{subfig:2e} and \ref{subfig:2f} that
the latter approach results in a much sharper image from which the
real space lattice points can be easily extracted. 

To quantify this
difference we normalized recovered real space intensity values of both
methods and sorted them in decreasing order. We plotted the sorted
values as 1D curves in Figure \ref{fig:3}. The curve that separates
the orange and blue region of the plot is associated with the solution to
the L1 minimization problem.  The curve that separates the blue region
and the white area above it is associated with the sorted intensities
obtained from a direct 3D inverse FFT. Clearly, the intensity
associated with the solution to the L1 minimization problem decreases
much more rapidly, thereby making it easy to select a threshold
(shown as the magenta line in Figure~\ref{fig:3}) that can be used to
identify real space lattice points.

\section{Concluding Remarks}
We presented a new technique for autoindexing nanocrystal diffraction
images.  The technique is based on dividing the indexing problem in
three steps. We reformulate the critical step of recovering a real
space 3D map of the lattice as an L1 minimization (or BP) problem and
solving the problem by an efficient and robust numerical algorithm.  A
simple numerical thresholding reveals the positions of the 3D lattice
points in real space. They can subsequently be used to determine the
unit cell parameters, crystal orientation and type as currently done 
in existing crystallographic software. Mirror symmetries of the lattice
generate multiple solutions that still need to be sorted out in a
final step using the measured intensities.

 Although the algorithm is more costly than the existing
approach because it is iterative and performs multiple 3D FFTs, it has
the advantage of recovering crystal lattice reliably when only a few
Bragg peaks can be measured.  Greedy algorithms that make use of the
sparsity of the solution may avoid this problem and increase speed
significantly \cite{omp,romp,cosamp,stomp,block-sparse}. 

Once the lattice vectors and orientation are determined, the
  lattice coordinates that overlap with the Ewald sphere will provide
  the index of a reflection.

 We demonstrate the feasibility of the technique with a
simple example.  More studies are needed to test the efficacy of the
method on different types of Bravais lattices and on datasets that may
be contaminated with noise.  However, we believe that our preliminary
results already indicate that compressive sensing based autoindexing
is a promising tool for ultrafast nanocrystallography.  Moreover, this
type of technique allows other constraints to be easily incorporated
into L1 minimization formulation to improve the reliability of
indexing.  It may even be possible to extend this approach to index
multiple crystals, powder diffraction or Laue data.

\noindent {\bf Acknowledgment}

We thank J. C. H. Spence for raising the problem, and Y. C. Eldar for
discussions.  This work was supported by the Laboratory Directed
Research and Development Program of Lawrence Berkeley National
Laboratory under U.S. Department of Energy Contract
No. DE-AC02-05CH11231.  Funding for FRNCM was provided by the American
Recovery and Reinvestment Act Computational Science and Engineering
Petascale Initiative.

\sectionmark{Bibliography}
\bibliographystyle{model1-num-names}
\bibliography{csindexing_arxiv}

\end{document}